\documentclass{mem}
\usepackage{natbib}
\usepackage{txfonts}
\usepackage{balance}
\usepackage{graphicx}
\usepackage[a4paper]{hyperref}
\idline{10}{10}
\begin{document}
\def\teff{$T\rm_{eff }$}
\def\kms{$\mathrm {km s}^{-1}$}

\title{
New high resolution synthetic stellar libraries  for the Gaia Mission
}

   \subtitle{}

\author{
R. \,Sordo,\inst{1}, 
A. \,Vallenari\inst{1},
J.-C. \,Bouret  	\inst{2},	
I. \,Brott     		\inst{3},	
B. \,Edvardsson  	\inst{4},
Y. \,Fr\'emat  		\inst{5},
U.~\,Heber    		\inst{6},
E. \,Josselin  		\inst{7},
O. \,Kochukhov  	\inst{8},
A. \,Korn 		\inst{8},	
A. \,Lanzafame  	\inst{9},
F. \,Martins            \inst{7}	
A.~\,Schweitzer  	\inst{10}, 
F. \,Th\'{e}venin	\inst{11}
J. \,Zorec		\inst{12}
}
\offprints{R. Sordo\\ \email{rosanna.sordo@oapd.inaf.it}}


\institute{
INAF--Osservatorio Astronomico, Vic. dell'Osservatorio 5
I-35122 Padova, Italy
\and
Laboratoire d'Astrophysique de Marseille, CNRS-Universit\'e de Provence, BP 8, 13376 Marseille Cedex 12, France
\and
Sterrenkundig Instituut, Princetonplein 5, NL-3584 CC Utrecht, The Netherlands
\and
Department of Natural Sciences, Mid Sweden University, S-851 70 Sundsvall, Sweden
\and
Royal Observatory of Belgium, 3 Avenue Circulaire, 1180 Brussels, Belgium
\and
Dr. Remeis-Sternwarte, Astronomisches Institut der Universit\"at Erlangen-N\"urnberg, Sternwartstr. 7, 96049 Bamberg, Germany
\and
GRAAL, Universit\`e Montpellier II - ISTEEM, CNRS, Place Eugene Bataillon, 34095 Montpellier Cedex, France
\and
Dept. of Physics and Astronomy, Uppsala University, BOX 515, 75120 Uppsala, Sweden
\and
Dip. di Fisica e Astronomia, Universit\`a di Catania, Via S. Sofia 78, 95123 Catania, Italy
\and
Hamburger Sternwarte, Gojenbergsweg 112, 21029 Hamburg, Germany
\and
O.C.A., UMR 6202, BP 4229, F-06304, Nice Cedex 4, France
\and
Institut d'Astrophysique de Paris, UMR7095 CNRS, Universit\'e Pierre\,\&\,Marie Curie, 98 bd. Arago, 75014 Paris, France}

\authorrunning{Sordo, Vallenari et al.}

\titlerunning{Stellar libraries for Gaia}

\abstract{
High resolution synthetic stellar libraries are of fundamental importance for the preparation of the Gaia Mission. We present new sets of spectral stellar libraries  covering two spectral ranges: 300 --1100 nm at 0.1 nm resolution, and 840 -- 890 nm at 0.001 nm resolution. These libraries span a large range in atmospheric parameters, from super-metal-rich to very metal-poor (-5.0 $<$[Fe/H]$<$+1.0), from  cool to hot (\teff=3000--50000 K) stars, including peculiar abundance variations.
The spectral resolution, spectral type coverage and number of models represent a substantial improvement over previous libraries used in population synthesis models and in atmospheric analysis.

}
\maketitle{}

\section{Introduction}

\begin{table*}[th!]
\scriptsize
\caption{Synthetic stellar libraries for Gaia. References: $^1$\cite{Bouret08},$^2$\cite{Hind}, $^3$\cite{Koch05}, $^4$\cite{Gust08}, $^5$\cite{Alva98}, $^6$\cite{Brott05}, $^7$\cite{WD}}
\label{sinopt} 
\begin{center}
\vspace{-0.3cm}
\begin{tabular}{l l l l l}
\hline 
\\ [-3pt]
Name &  T$_{eff}$  (K)& $\log g$ & [Fe/H]  &Notes \\[3pt]
\hline
\\
O,B,A stars $^1$  & 8000 -- 50000 & 1.0 -- 5.0&-5.0 -- +1.0&TLUSTY code, NLTE, wind, mass loss		\\[2pt]
sdO   $^2$             &26000 -- 100000 & 4.8-6.4  & +0.0    0&TMAP, NLTE, $\Delta$\teff= 500 K, $\Delta \log g$= 0.2 \\[2pt]
A stars	$^3$      & 6000 -- 16000 & 1.0 -- 5.0&+0.0        &Magnetic field	\\[2pt]
MARCS	$^4$      & 4000 -- 8000  &-0.5 -- 5.0&-5.0 -- +1.0& Variations in individual $\alpha$-elements abundances	\\[2pt]
C stars	$^{4,5}$  & 4000 -- 8000  & 0.0 -- 5.0&-5.0 -- +0.0& $\Delta$\teff= 500 K; [C/Fe]=0,1,2,3; [$\alpha$/Fe]=+0.0, +0.4		\\[2pt]
PHOENIX	$^6$      & 3000 -- 10000 &-0.5 -- 5.5&-3.5 -- +0.5& $\Delta$\teff= 100 K; [$\alpha$/Fe]=-0.2--+0.8, $\Delta$[$\alpha$/Fe]=0.2\\[2pt]
Emission lines    &$\geq$ 15000 & 2.8 -- 4.0  &+0.0        &Be, WR models                                  \\[2pt]
WD$^7$            & 6000 -- 90000 &7.0--9.0     &            &WDA \& WDB, LTE					\\[2pt]
\hline
\end{tabular}
\end{center}
\vspace{-0.4cm}
\end{table*}
\normalsize

ESA's Gaia mission, to be launched in 2011, is meant to obtain accurate position, parallax and  proper motion for 10$^9$ object all over the sky, up to magnitude $G$=20 ($V$=20--22) with an astrometric accuracy at the $\mu$ arcsec level, providing low-dispersion spectroscopy for each object and radial velocity up to $G$=16 (for more detailss see Cacciari 2008, this meeting).
The final catalogue is expected to provide the discrete classification of sources (single stars, galaxies, QSOs, asteroids) and the astrophysical parameters (APs) for single stars (i.e. T$_{\rm eff}$, $\log g$,...) and possibly the parametrization of special sources (galaxies).
Gaia will produce an unprecedented amount of data (40--50~GB/day) leading to 100~TB of compressed data in 5 years. This huge number of observed objects must be classified in an automated way. 
In the Gaia project, all the classification algorithms  but one are based on supervised methods  \citep[see for example][]{Coryn07}, classifying a source or estimating its APs by comparison with a set of templates.
In order to do this,  simulations of Gaia observations covering the whole AP space are required: high resolution and high quality synthetic libraries are of fundamental importance. Here we focus on new stellar libraries, while galaxies, QSOs, asteroids are discussed elsewere \citep{Vivi, Claeskens, Warell}.

\section{New stellar libraries}
\begin{figure*}[th!]
\begin{center}
 \resizebox{6.7truecm}{!}{\includegraphics[clip=true,angle=0]{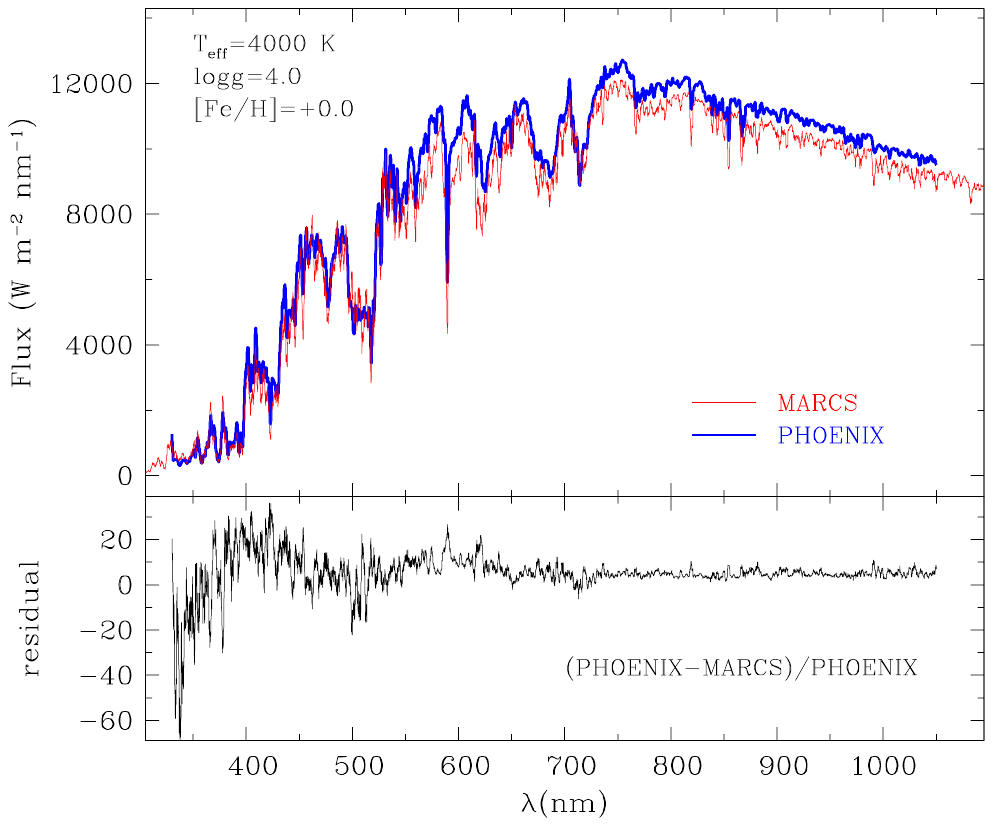}}
\resizebox{6.7truecm}{!}{\includegraphics[clip=true,angle=0]{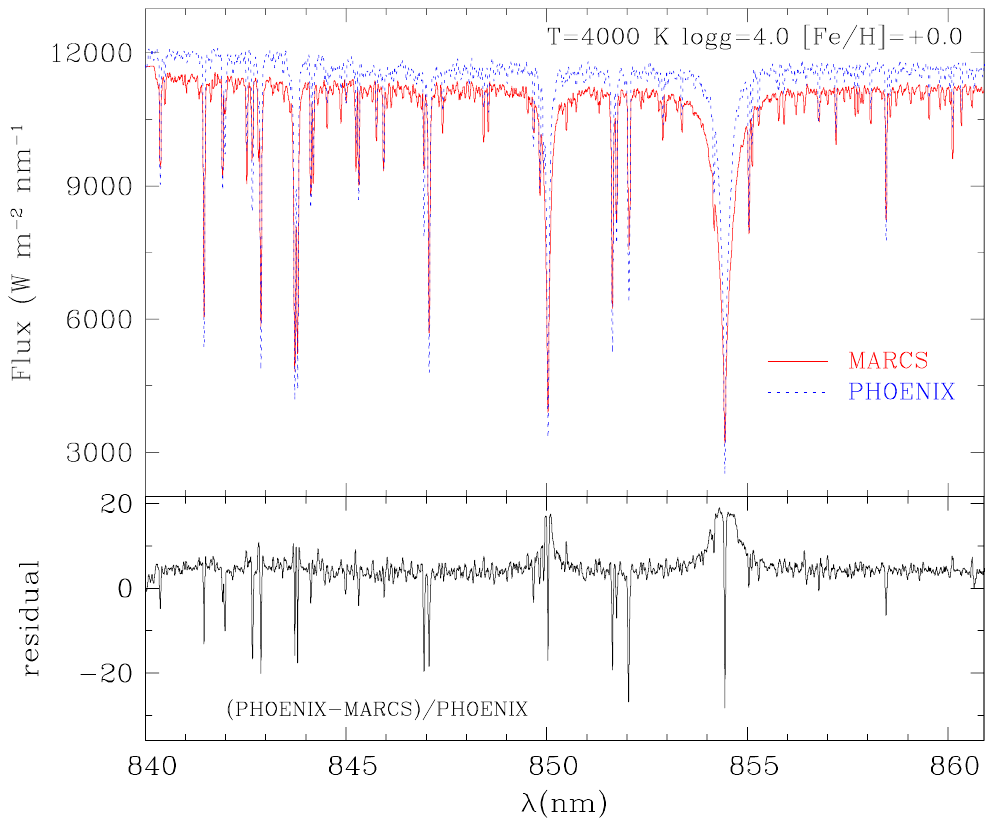}}
 \caption{
Comparison between two different libraries (MARCS, PHOENIX) for a star having \teff=4000 K, $\log g$=4.0, [Fe/H]=+0.0, in the BP/RP range (left panel) and in a selected region of the RVS range (right panel). Residuals (PHOENIX - MARCS)/PHOENIX are also shown in the lower panels.
}
 \label{low_R}
\end{center}
 \end{figure*}

A large community of scientists has agreed to produce state-of-the-art libraries of synthetic spectra, with an homogeneous and complete coverage of the AP space at the two fixed resolutions required to produce Gaia simulations: 0.1 nm for the low-dispersion (300--1100 nm) and 0.001 nm for the high resolution mode (840--890 nm).
 The capability of reproducing real spectra is improving, and each code producing synthetic spectra is tuned for a given type of stars.  These new libraries, summarized in Table.~\ref{sinopt}, span a large range in atmospheric parameters, from super-metal-rich to very metal-poor stars, from cool stars to hot, from dwarfs to giant stars, with small steps in all parameters, typically $\Delta$\teff=250~K (for cool stars), $\Delta \log g$=0.5 dex, $\Delta$[Fe/H]=0.5 dex.  Depending on \teff, these libraries rely on MARCS (F,G,K stars), PHOENIX (cool and C stars), KURUCZ, TLUSTY (A,B,O stars) models. Those models are based on different assumptions: KURUCZ are LTE, plane-parallel, MARCS implements also spherical symmetry while PHOENIX and TLUSTY (hot stars) can calculate NLTE models both in plane-parallel mode and spherical symmetry \citep[see for a more detailed discussion][]{Gust08}. 
MARCS spectra are also calculated including a global $\alpha$ enhancement (from -0.2 to 0.4 with a step of 0.2 dex). Moreover, enhancements of individual $\alpha$ elements (O, Mg, Si, Ca) are considered.
Hot star spectra  take into account the effect of magnetic field, peculiar abundances,  mass loss, and circumstellar envelope (Be).
In the next future, the calculation of stellar libraries in the very low temperature regime is foreseen.

The large overlap in the AP space among different libraries will allow their comparison. 
A large effort is ongoing in literature to compare spectra of different stellar libraries \citep[see][]{Gust08, Martins07}.

 \begin{figure*}[th!]
\begin{center}
 \resizebox{6.47truecm}{!}{\includegraphics[clip=true,angle=0]{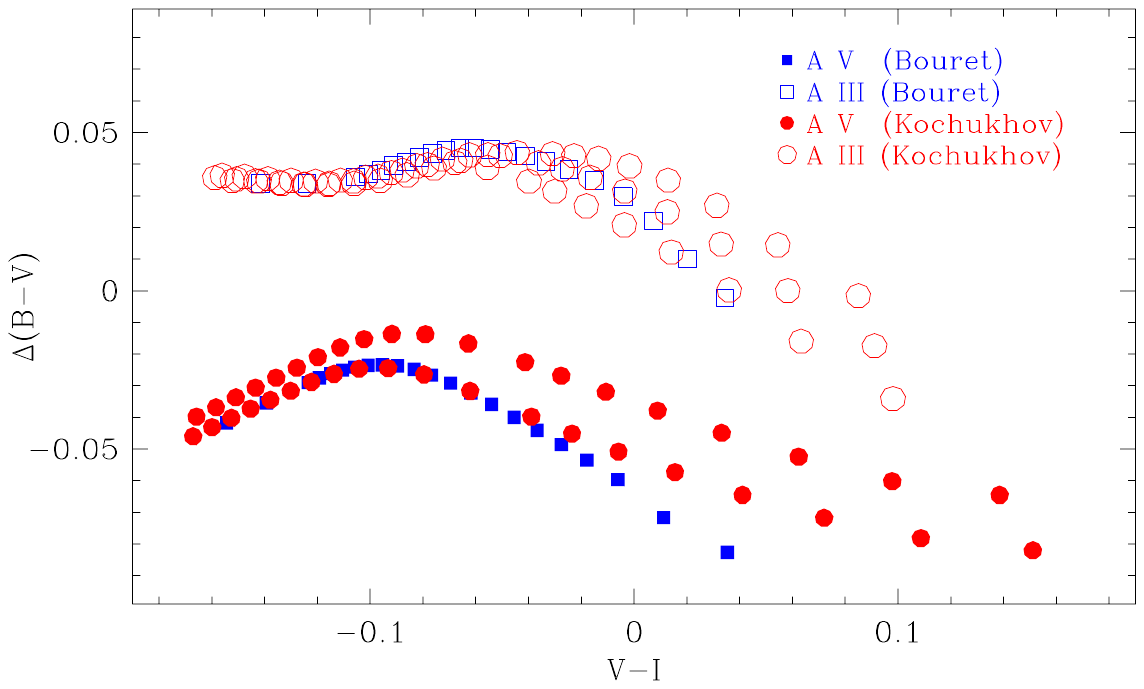}}
 \resizebox{6.47truecm}{!}{\includegraphics[clip=true,angle=0]{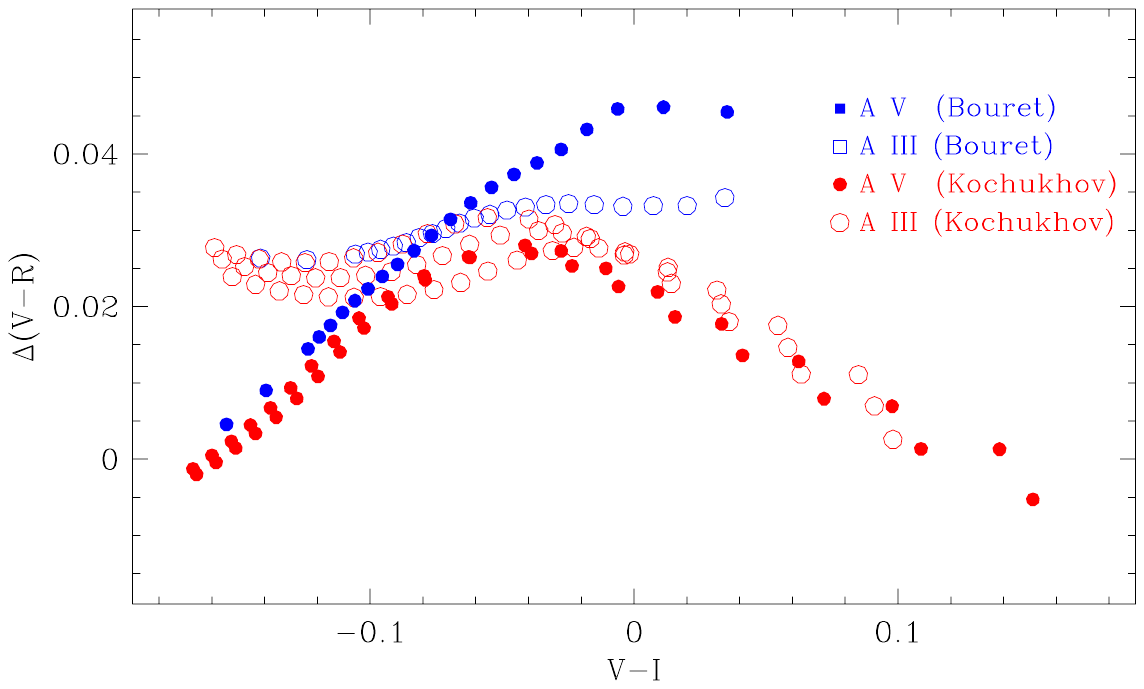}}\\
 \resizebox{6.47truecm}{!}{\includegraphics[clip=true,angle=0]{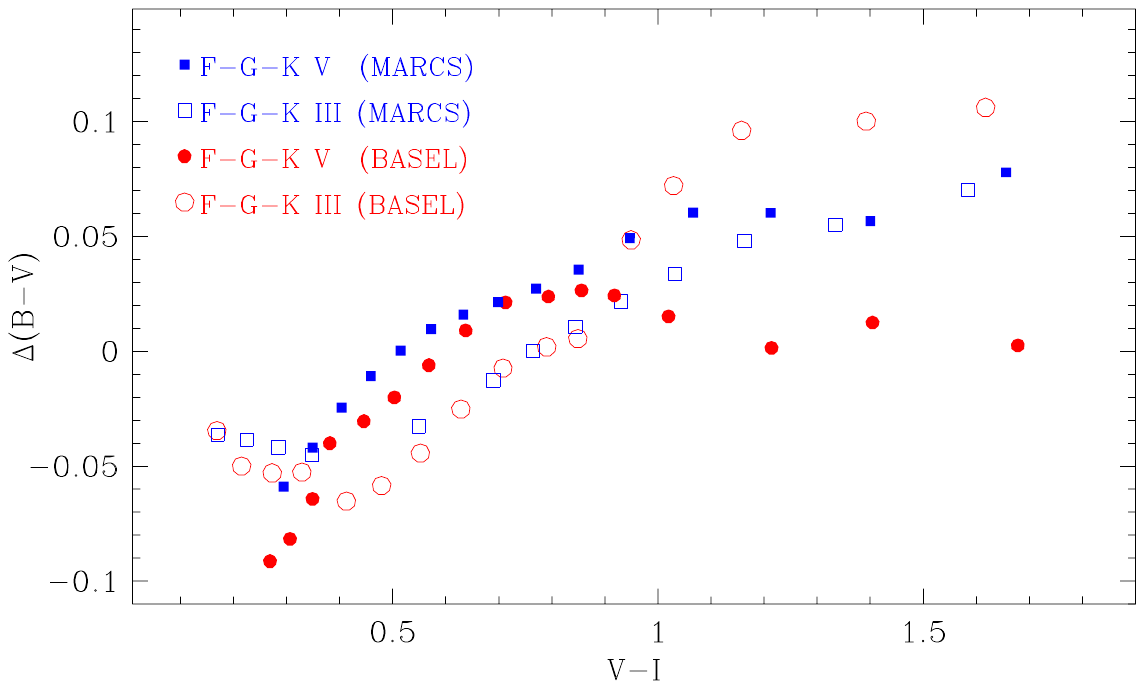}}
 \resizebox{6.47truecm}{!}{\includegraphics[clip=true,angle=0]{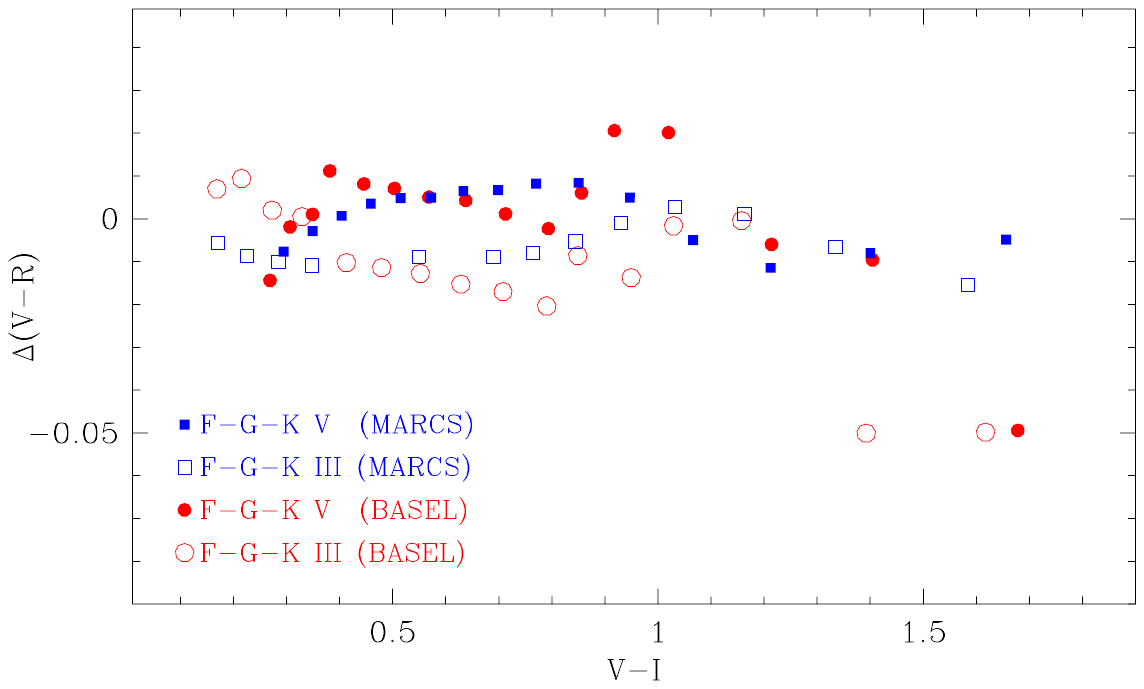}}
 \caption{
Comparison in the (V-R)-(V-I) diagram among all available high resolution libraries covering the 8000--15000 K T$_{\rm eff}$ range covered by A stars, for solar metallicity. Bottom panel shows difference between each library and the empirical calibration of \cite{Worthey06}.
 }
\label{A_VR}
\end{center}
 \end{figure*}


The  impact of the underlying assumptions, of the different input physics (i.e. atomic and molecular line lists, convection treatment) or of the inclusion of NLTE effects, is shown in Fig.~\ref{low_R}, where MARCS and PHOENIX are compared in the case of \teff=4000 K, $\log g$=4.5, solar metallicity. 
In general, differences between the two libraries are of the order of 10\% -20\% in the blue part of the spectrum, being worse for low temperature stars, as expected.  Hot stars (O, B, A) are in  agreement at the 1-2\% level. Broad band colors (B-V, V-R, V-I) are good tracers of the flux distribution given by different libraries.  Fig.~\ref{A_VR} shows the comparison of some libraries  at solar metallicity, for dwarfs and giant stars with the empirical calibration by \cite{Worthey06}. \\ In  Fig.~\ref{A_VR} (upper panels) two sets of spectra of A stars are compared: A (Bouret), A (Kochukhov). They show a similar behaviour and are a good reproduction of the empirical relations, in the diagram (V-R)-(V-I), where the residuals are  $< 0.05 $ mag. The agreement is  worse in the (B-V)-(V-I) diagram.\\
 In Fig.~\ref{A_VR} (lower panels) two sets of spectra of F-G-K stars are compared: MARCS and BASEL \citep{Lejeune}. These libraries show a comparable agreement with the data. In the diagram (V-R)-(V-I), the residuals are $<$ 0.07 mag, while in the (B-V)-(V-I) the residuals are of the order of $\pm 0.1$.

\section{Conclusions}

In this paper we summarize the work recently done by a large scientific comunity in the calculation of  a large database of state-of-the-art synthetic stellar spectra, to be used in preparation of the ESA's Gaia mission. They cover the whole optical range with 0.1 resolution and the Ca II triplet region at 0.001 nm resolution, for a large set of astrophysical parameters. Generally, these libraries are of great interest in stellar population synthesis and abundance analysis works.
These large grids offer excellent possibilities to explore the variation of the spectrum properties with stellar parameters.  Even if  theoretical stellar spectra modelling has greatly improved in the paste decades, still the effects  and the inconsistency  introduced by the underlying simplifying assumptions need to be analysed in detail. Here   a few comparison are shown, using broad band colors and direct spectra comparison. While (V-I) and (V-R) color are a good reproduction of the empirical calibration for MARCS, Basel, A stars, (B-V) colors give a poorer result, being greatly affected by the well known problem of the  line list. Direct spectra comparison shows that hot star spectra are in reasonable agreement. The differences between MARCS and PHOENIX are of the order of 10-20\%, in particular for cool stars and at the bluest end of the spectrum.

\begin{acknowledgements}
This short review relies on the work of many members of the Gaia community, who are gratefully anknowledged for their valuable contribution to the project. 
\end{acknowledgements}

\bibliographystyle{aa}

\end{document}